# Supercooled-liquid and plastic-crystalline state in succinonitrile-glutaronitrile mixtures

M. Götz, Th. Bauer, P. Lunkenheimer,[a] and A. Loidl

*Experimental Physics V, Center for Electronic Correlations and Magnetism, University of Augsburg, 86135 Augsburg, Germany*

We report a thorough characterization of the glassy phases of mixtures of succinonitrile and glutaronitrile via dielectric spectroscopy and differential scanning calorimetry. This system is revealed to be one of the rare examples where both glassy states of matter, a structurally disordered supercooled liquid and an orientationally disordered plastic crystal, can be prepared in the same material. Both disordered states can be easily supercooled, finally arriving at a structural-glass or a glassy-crystal state. Detailed investigations using broadband dielectric spectroscopy enable a comparison of the glassy dynamics in both phases. Just as previously demonstrated for supercooled-liquid and plastic-crystalline ethanol, our experiments reveal very similar relaxational behavior and glass temperatures of both disordered states. Thus the prominent role of orientational degrees of freedom in the glass transition, suggested on the basis of the findings for ethanol, is fully corroborated by the present work. Moreover, the fragilities of both phases are determined and compared for different mixtures. The findings can be well understood within an energy-landscape based explanation of fragility.

## I. INTRODUCTION

The glass transition is a prominent unresolved problem of the physics of condensed matter, despite, on an empirical level, it is used partly since centuries in numerous technical applications.[1,2,3,4] In its original sense, the glass transition as employed, e.g., by glass blowers means a continuous increase of viscosity $\eta$ under cooling, enabling to finely tune $\eta$ to a desired level, suitable for casting, blowing, etc. On a microscopic level, this variation of the viscosity implies a continuous freezing of *translational* motions of the structural units (which, depending of the class of the glass former, can be molecules, metal ions, polymer segments, etc.). Understanding this slowing down is essential for understanding the glass transition in general. This glassy dynamics very often is investigated by dielectric spectroscopy.[5,6,7] Due to the very broad dynamic range accessible with this method, it is possible to follow the many-decades variation of the molecular time scale from very fast molecular motions in the liquid (ps range) to the total arrest below the glass temperature $T_g$ (>100 s). Curiously, dielectric spectroscopy does not provide direct access to the *translational* motions, whose freezing drives the glass transition, but instead probes *reorientational* dynamics only, at least in its most common version where non-ionic, dipolar molecular glass formers are investigated. Fortunately, in most cases sufficiently good coupling of both degrees of freedom can be stated. While minor or major decoupling effects are known to occur, depending on the system (see, e.g., refs. 1,2,8), in most cases the reorientational dynamics accessed by dielectric spectroscopy enables meaningful conclusions on the glass transition. A prominent example is the supercooled liquid (SL) glycerol, which is one of the most investigated glass formers: There a good coupling of both degrees of freedom can be rationalized by considering that the ubiquitous hydrogen bonds existing between molecules have to break for translational as well as for reorientational motions.

However, there is a class of glasslike systems where *complete* decoupling of both dynamic modes occurs, the so-called plastic crystals (PCs).[9,10] In these materials only the orientational degrees of freedom of the molecules show glassy freezing while their centers of mass are completely fixed on a crystalline lattice with strict translational symmetry, thus showing no significant translational diffusion at all. The investigation of PCs formed by dipolar molecules with dielectric spectroscopy reveals reorientational dynamics that shows all the phenomenology of "normal" glass formers as, e.g., the non-Arrhenius slowing down of the molecular relaxation time and non-exponential relaxation.[10] Thus these materials are considered as simpler model systems for "true" glass formers, especially if considering that the reorientational dynamics often is found to show a glass transition, leading to a so-called glassy crystal.[11]

Interestingly, there are a few cases where the occurrence of both glassy phases in one and the same substance was reported, which can be prepared depending on cooling/heating history.[12,13,14,15,16,17] This provides the unique opportunity to investigate the role of the orientational degrees of freedom in the glass transition by direct comparison of the glassy dynamics in both disordered states for the same type of molecule. So far, only one case, ethanol, has been investigated in detail in both phases and astonishing similarities of the "true" supercooled and plastic crystalline state were found.[10,18,19,20,21] For example, the general shape of the dielectric spectra is similar, the relaxation times characterizing molecular motion are of similar order of magnitude, and the glass transition occurs at nearly the same temperature. These studies led to the unexpected conclusion that the glass transition, at least in ethanol, to a large extent is governed by the freezing of the orientational degrees of freedom! This finding is surprising because usually the glass transition is believed to be triggered by the continuous freezing of translational motions, which leads to the well-known increase of viscosity. Is this behavior a unique property of ethanol or is the glass transition in other glass formers also much more governed by orientational motions than previously thought? In this context, one should have in mind that many monohydroxy alcohols are known to exhibit unusual relaxation dynamics: There the main relaxation process is not the structural $\alpha$ relaxation determining viscous flow but most likely due to correlated motions of clustered molecules bound to each other via hydrogen bonds.[22] While the situation for ethanol is not so clear,[19] overall the relaxation dynamics of monohydroxy alcohols seems to be a special case and the results found for ethanol could be suspected to be of limited general importance for the glass transition.

[a] Electronic mail: peter.lunkenheimer@physik.uni-augsburg.de

For these reasons, it would be desirable to check for the universality of the findings in ethanol by investigating additional materials that can be prepared in both disordered states. Unfortunately, in the few other materials known to show a plastic-crystalline *and* a SL state, the latter state is difficult to access, requiring, e.g., fast quenching or even hyperquenching that is possible by molecular dynamics simulations only.[17] Consequently, the dynamics of none of these materials has been investigated as nearly as thoroughly as ethanol. However, in the present work we show that there is another system where both glassy states can be easily accessed, namely mixtures of succinonitrile and glutaronitrile. In this system, until now only the plastic crystalline state of a mixture of 60% succinonitrile and 40% glutaronitrile was thoroughly investigated.[23] In addition, a phase diagram was published revealing the presence of PC and glassy crystal phases for a large range of concentrations.[23] In the present work, we show that in a certain concentration range sufficiently fast cooling of these mixtures can also produce a supercooled state. We find an astonishing agreement of the properties of both disordered states in this system with the general behavior previously found for supercooled and plastic-crystalline ethanol. Thus it seems that the importance of orientational degrees of freedom for the glass transition indeed may be universal.

## II. EXPERIMENTAL DETAILS

Succinonitrile and glutaronitrile with stated purities of $\geq$ 99% were purchased from Arcos Organics and measured without further purification. The mixtures were prepared by putting liquid glutaronitrile into succinonitrile, melted in a water bath, under heavy stirring. The concentrations are specified in mol%. To check for phase transitions and glass anomalies, the sample materials were characterized by differential scanning calorimetry (DSC) using different heating rates.

Two experimental techniques were combined to arrive at dielectric spectra of the complex permittivity covering a frequency range of about $10^{-1}$ Hz - 3 GHz.[24] In the low-frequency range, $\nu <$ 3 MHz, a frequency-response analyzer (Novocontrol $\alpha$-analyzer) was used. Measurements in the radio-frequency and microwave range (1 MHz < $\nu$ < 3 GHz) were performed using a reflectometric technique where the sample capacitor is mounted at the end of a coaxial line.[25] For these measurements an Agilent E4991A impedance analyzer was employed. For both methods, the sample material was filled into parallel-plate capacitors with plate distances betwen 50 and 150 µm. For cooling and heating of the samples, a nitrogen gas cryostat (Novocontrol Quatro) was used.

## III. RESULTS AND DISCUSSION

### A. DSC and phase diagram

Figure 1 shows DSC results for the mixture of 20% succinonitrile and 80% glutaronitrile (20SN-80GN), obtained for different cooling/heating rates. When cooling the liquid mixture with a relatively low rate of 2 K/min [Fig. 1(a)], a clear exothermic minimum at about 181 K indicates crystallization. Under further cooling a small steplike anomaly shows up at about 150 K (see inset), which is more clearly pronounced at heating. Its shape is typical for a glass transition, proving that the liquid has transformed into a plastic-crystalline state below 181 K, which finally becomes a glassy crystal below the orientational glass temperature $T_g^o \approx 151$ K. Similar results for other mixing ratios have led to a phase diagram showing that the plastic crystalline and glassy crystal state occurs in a broad range of about 15 - 96% glutaronitrile.[23]

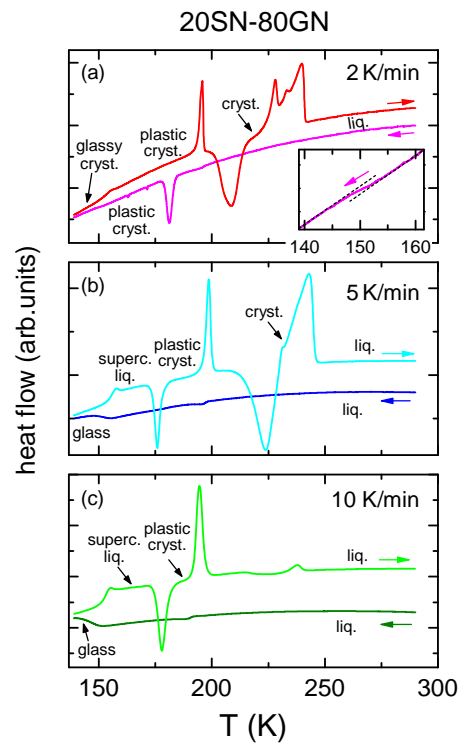

FIG. 1. DSC results for 20SN-80GN for different cooling/heating rates. The inset shows a zoomed view of the cooling curve in (a) at low temperatures. The dashed lines illustrate the presence of a small anomaly at about 150 K.

Interestingly, in contrast to the findings for a moderate rate of 2 K/min, at cooling rates of 5 and 10 K/min the crystallization minimum of 20SN-80GN is completely missing and only a glass transition, much better pronounced than for 2 K/min, is observed [Fig. 1(b) and (c)]. This result clearly proves that at these higher rates crystallization can be avoided and the liquid becomes supercooled and finally forms a "true" or structural glass below $T_g \approx 150$ K. Obviously, similar to ethanol, both disordered phases are accessible in this glass former and the glass transition temperatures are nearly identical. When heating beyond the glass-transition temperature, for cooling/heating rates of 5 and 10 K/min [Fig. 1(b) and (c)], spontaneous crystallization into the plastic-crystalline phase occurs. This is revealed by the sharp minima at about 177 K. This phase melts at about 195 - 198 K as indicated by the endothermic maximum at this temperature. The same melting occurs for the sample cooled with 2 K/min [Fig. 1(a)], which is in the plastic-crystalline phase already above $T_g^o$.

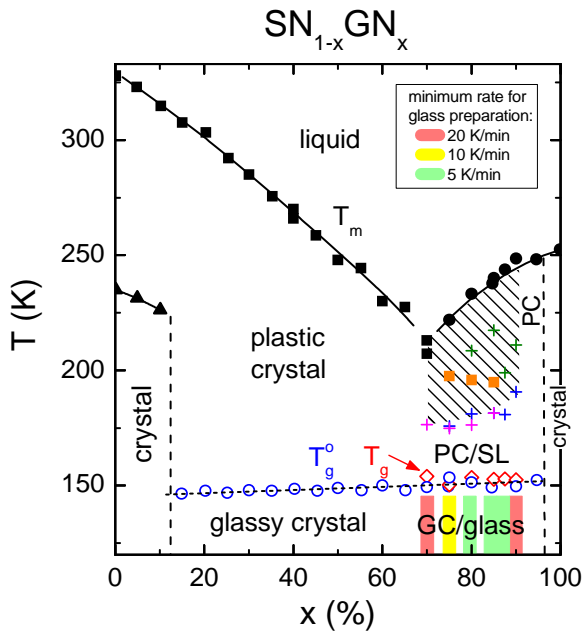

FIG. 2. Phase diagram of the system $SN_{1-x}GN_x$ as determined from DSC measurements. The diagram is based on that published in ref. 23. The open circles show the known orientational glass temperatures from the PC to the glassy crystal (GC).[23] The conventional glass temperatures, determined in the present work, which mark the transition from the SL phase into the glass state, are indicated by the open diamonds. Colored bars indicate the region where a structural glass state can be formed for cooling rates ≤ 20 K/min; the colors indicate the necessary cooling rates (see figure legend). In the hatched area, the behavior is complex and depends on history and cooling/heating rate. Closed squares indicate the melting transition of the plastic crystalline phase under heating. The closed circles denote the final melting into the liquid when recrystallization has occurred: most probably this is the melting transition of the completely ordered crystal. The crosses show spontaneous crystallization transitions from the SL into the PC under heating (green and magenta) or cooling (blue). The triangles at low $x$ indicate the transitions from the completely ordered (i.e., both translationally and orientationally) to the plastic-crystalline state.[23]

For the 10 K/min measurement, heating beyond this melting peak reveals no further significant anomalies, except for a tiny peak at 238 K, which may arise from some residual phase. However, for the 2 and 5 K/min samples the situation above about 200 K is more complex: At temperatures beyond the melting peak of the PC, another exothermic transition shows up at about 210 (2 K/min) and 225 K (5 K/min) indicating the transition into another crystalline phase. This phase then finally melts at 230 - 240 K in a succession of transitions (2 K/min) or a single smeared-out transition (5 K/min). The nature of these transitions occurring under heating after the PC has melted is not clear and one may speculate that phase separation may play some role here. However, interestingly quite similar behavior was also found in ethanol: There the melting peak of the PC was followed by a recrystallization minimum, most likely marking the transition into the completely ordered crystal, which finally melted at even higher temperatures.[12] A similar scenario may also explain the present results, i.e. the highest endothermic peak observed in Figs. 1(a) and (b) is due to the final melting of the completely ordered crystal into the liquid.

Several further samples with different mixing ratios were measured using DSC and checked for their ability to form a SL and structural glass. The results are indicated in the phase diagram, Fig. 2, which is based on the diagram published in ref. 23. As indicated by the colored bars, a structural glass state can be best achieved at glutaronitrile concentrations around 80%, where relatively moderate cooling rates of 5 K/min are sufficient to avoid crystallization. The found glass temperatures and other transition temperatures are indicated in the figure. As already discussed for 20SN-80GN, also for the other mixtures that form a glass, $T_g$ and $T_g^o$ are virtually identical. In the shaded region, above the crystallization temperature into the PC (magenta and blue crosses), complex behavior arises and only the melting of the PC (orange squares) and the final melting after recrystallization (closed circles) can be considered as well-established.

## B. Dielectric spectra

Based on the DSC results and the obtained phase diagram (Fig. 2), dielectric spectra were collected in the two disordered states of the samples. In the following, dielectric spectra will only be provided for sample preparations and temperature regions where the phase assignment is clear, based on the performed DSC measurements. Figure 3 shows the results for 20SN-80GN in the liquid (closed symbols) and in the plastic-crystalline or glassy-crystal state (open symbols). The data were obtained in a single measurement run under cooling. A moderate rate of 0.4 K/min was chosen and the phase behavior of the sample can be assumed to be similar to that shown by the cooling curve in Fig. 1(a), i.e. a transition from the liquid/SL to the PC is expected at about 180 K (cf. blue cross at $x = 80\%$ in Fig. 2). In Fig. 3, typical relaxational response is observed as revealed by a steplike decrease of the real part of the permittivity $\varepsilon'(\nu)$ with increasing frequency and a corresponding peak in the dielectric loss $\varepsilon''(\nu)$.[26] In both phases, this behavior can be ascribed to the reorientational motions of the molecules, i.e. it corresponds to the $\alpha$ relaxation as commonly found in dipolar PCs and glass form-ing liquids.[5,6,7,10] Both spectral features strongly and continuously shift to lower frequencies when the temperature is lowered, mirroring the glassy freezing of these motions. For temperatures coming close to the orientational glass transition at $T_g^o \approx 151$ K (cf. Figs. 1 and 2) and in the glassy-crystal state at $T < T_g^o$, these main relaxation features have shifted out of the frequency window.

Already a simple inspection by eye of the spectra of Fig. 3 reveals that the transition from the translationally and reorientationally disordered liquid to the solely reorientationally disordered plastic crystalline state does not lead to any dramatically different relaxational behavior: Both the shape and the amplitude of the observed relaxational features show a rather smooth temperature development without any obvious change at the transition occurring at about 181 K. However, the somewhat larger frequency distance of the steps and peaks at 180 and 186 K, compared to the neighboring spectra, indicates a small jump in the relaxation time $\tau$, which can be estimated from the peak frequency $\nu_p$ via $\tau = 1/(2\pi\nu_p)$. The relaxation-time development will be treated in more detail below.

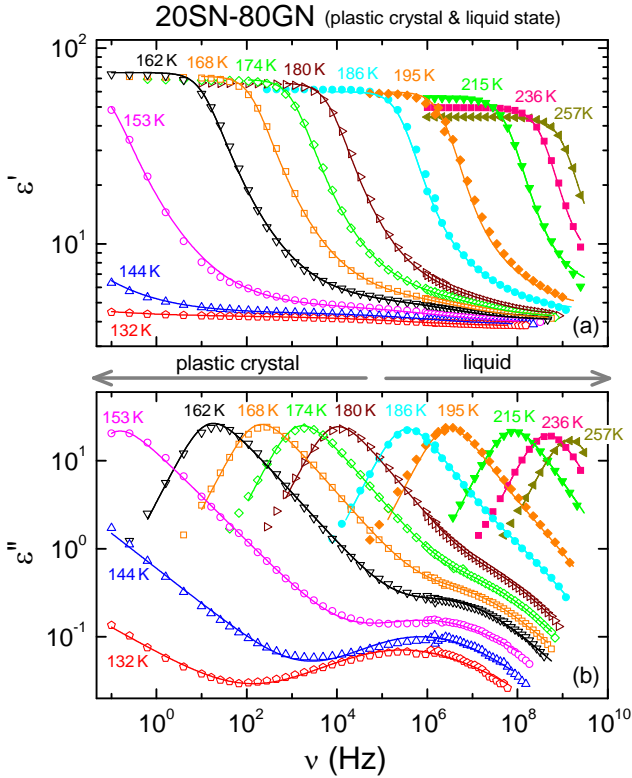

FIG. 3. Spectra of the dielectric constant (a) and loss (b) of 20SN-80GN, measured at various temperatures. The spectra have been obtained in a single measurement-run under cooling. The closed symbols show results in the liquid state. The spectra indicated by the open symbols were measured in the plastic-crystalline ($T > 150$ K) or glassy-crystal state ($T < 150$ K). The lines are fits with a CD function ($T \geq 215$ K) or the sum of a CD and a CC function ($T \leq 195$ K), simultaneously performed for $\varepsilon'$ and $\varepsilon''$.

In addition to the main relaxation, in the spectra of Fig. 3(b), for the lower temperatures a secondary relaxation peak of much smaller amplitude shows up at frequencies several decades faster than the main α-relaxation peak. Secondary relaxations, usually termed $\beta$ or $\gamma$ relaxations, are a common phenomenon in glass forming liquids[27,28,29] and are often also found in PCs (but usually with a rather weak amplitude only).[10,23,30] The secondary relaxation in Fig. 3(b) shows a weaker temperature-induced shift than the α relaxation and finally merges with the main relaxation at high temperatures, again a typical behavior.[28,29] Finally, based on the rather small slope of the $\varepsilon''(\nu)$ curve at 132 K and $\nu < 10$ Hz in Fig. 3, one may suspect the presence of another relaxation process between the $\alpha$ and the well-resolved secondary relaxation that is observed at about $10^6$ Hz for this temperature. A similar scenario was considered for 60SN-40GN in Ref. 23. Measurements at lower frequencies would be necessary to clarify this question, which, however, is out of the scope of the present work.

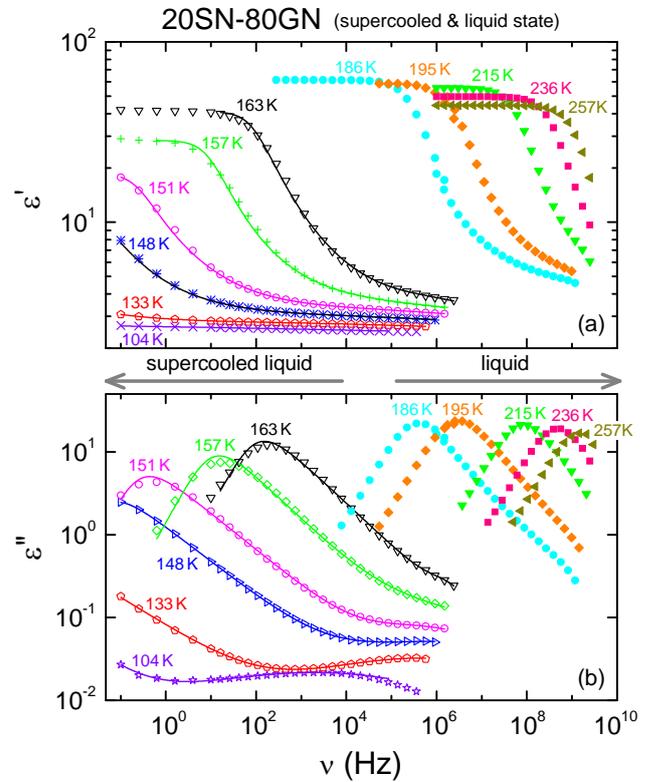

FIG. 4. Spectra of the dielectric constant (a) and loss (b) of 20SN-80GN, measured at various temperatures. The spectra at $T \leq 163$ K (open symbols) have been obtained under heating after quenching the sample and can be assigned to the supercooled-liquid and glass state. For comparison, also the spectra obtained in the liquid under cooling as already shown in Fig. 3 are included (closed symbols). The lines are fits with the sum of a CD and a CC function, simultaneously performed for $\varepsilon'$ and $\varepsilon''$.

Figure 4 shows the dielectric spectra for the same mixture (20SN-80GN), prepared in the SL state (open symbols). To thoroughly exclude crystallization, before the measurements the cryostat insert with the filled sample capacitor was quenched by immersing it into liquid nitrogen and afterwards inserted into the precooled cryostat. The measurements were performed under heating with 0.4 K/min. A succession of transitions as in Fig. 1(b) is found for this run. However, in Fig. 4 only data are shown that can be unequivocally assigned to the supercooled regime (cf. Fig. 2), i.e. only data taken at temperatures up to the point where spontaneous crystallization occurred. These data reveal that obviously a strong $\alpha$ relaxation also shows up in the supercooled state of this mixture. For comparable temperatures, the loss peaks and $\varepsilon'$ steps are located at somewhat higher frequencies than in the PC (cf. Fig. 3). The relaxation magnitude is somewhat smaller than in the plastic-crystalline state (Fig. 3). It decreases when the temperature is lowered, which is a rather unusual behavior, for which we currently have no explanation (successive crystallization can be excluded as these data were taken under heating). In Fig. 4(b), again a $\beta$-relaxation peak is found in the loss, which seems to be located at similar frequencies as in the plastic phase [cf. Fig. 3(b)]. For comparison, in Fig. 4 also the results in the liquid regime are provided by including the results at $T \geq 186$ K, already shown in Fig. 3 (closed symbols). The two data sets are consistent with a smooth development of the loss peaks and $\varepsilon'$ steps between the liquid and supercooled-liquid states. Only in $\varepsilon_\infty$, the high-frequency plateau value of $\varepsilon'(\nu)$, some discrepancy may be suspected as $\varepsilon_\infty$ seems to be somewhat lower in the supercooled regime. It is unlikely that $\varepsilon_\infty$, which is governed by the ionic and electronic polarizability, should really be different here. Thus, we ascribe this finding to an imperfect filling factor of the sample capacitor for the supercooled sample arising from the strong quenching performed before the measurement (the fast contraction of the sample material may lead to air entering the capacitor).

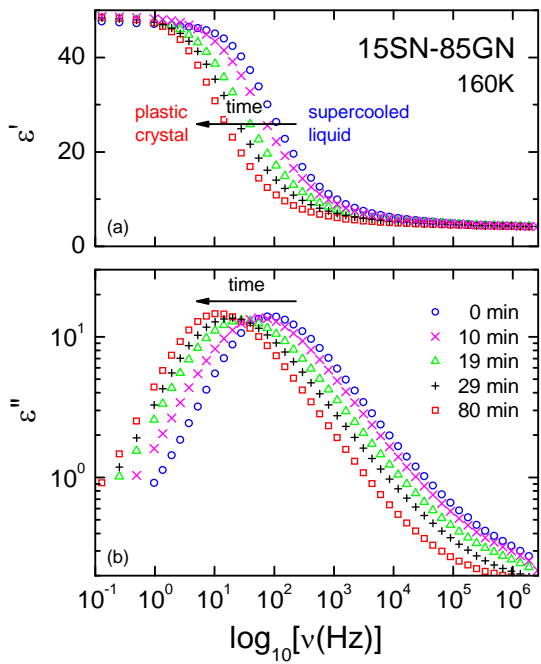

FIG. 5. Spectra of the dielectric constant (a) and loss (b) of 15SN-85GN, measured at 160 K at different times during the transition from the SL to the PC state.

Qualitatively similar dielectric behavior as shown in Figs. 3 and 4 was also found for the mixture of 15% succinonitrile and 85% glutaronitrile (15SN-85GN). In Fig. 5, we show dielectric spectra measured at different times during the transition from the supercooled liquid into the plastic-crystalline phase at 160 K. To obtain these results, the sample was first prepared in the supercooled state by rapid quenching. Afterwards the sample was heated to 160 K, where the transition into the more stable PC state took place. This temperature was carefully chosen to enable a smooth tracking of the transition; at slightly higher and lower temperatures, the transition rate was significantly faster or slower, respectively. In Fig. 5, it is nicely seen that this transition leads to a slowing down of the $\alpha$ relaxation, which is mirrored by a shift of the loss peak and $\varepsilon'$ step to lower frequencies. The final difference after the transition has completed is about one decade. The loss peak becomes broadened and reduced in amplitude during the course of the transition, which is ascribed to the simultaneous presence of both phases. The isosbestic point [$\varepsilon''(t) = $ const.] showing up in the loss curves in Fig. 5(b) is consistent with this scenario. A very similar behavior was also found for the corresponding transition between the two glassy phases in ethanol.[19]

The spectra shown in Figs. 3 and 4 were fitted by the sum of two relaxation functions. For the $\alpha$ relaxation, the empirical Cole-Davidson function was used,[31] which is often found to provide a good description of the main relaxation in various glass formers and plastic crystals.[6,7,10] The observed secondary relaxation features were fitted by the Cole-Cole function,[32] commonly used for fits of secondary relaxations.[10,28,29] The lines in Figs. 3 and 4 show the fit curves obtained from simultaneous fits of $\varepsilon'(\nu)$ and $\varepsilon''(\nu)$. A reasonable description of the experimental data is obtained in this way.

## C. Relaxation times

Fig. 6 shows the temperature dependence of the average relaxation times[33,34] deduced from the fits of the measured broadband dielectric spectra. Additional data points obtained from reading off the loss peak positions and using the relation $\tau = 1/(2\pi\nu_p)$ are also included. The closed symbols denote the results in the PC and glassy crystal phases, open symbols those obtained for the liquid, SL, and glass states. In general, the $\alpha$-relaxation times of the PC phase are somewhat slower than those of the supercooled-liquid phase. Such a moderate but significant difference of the relaxation times was also found for the two glassy phases of ethanol.[19,21] It can be rationalized by a stronger hindering of molecular reorientation in the more ordered and denser packed crystalline state.

For the $\alpha$ relaxation of both phases (circles), the Arrhenius representation of Fig. 6 reveals significant deviations from thermally activated behavior. The temperature dependence of the $\alpha$-relaxation times can be well described by the modified Vogel-Fulcher-Tammann law:[35,36]

$$\tau_\alpha = \tau_0 \exp\left[\frac{DT_{VF}}{T - T_{VF}}\right] \quad (1)$$

Here $D$ is the so-called strength parameter[36] and $T_{VF}$ the Vogel-Fulcher temperature; $\tau_0$ represents an inverse attempt frequency. As the available temperature range for the PC is rather limited and as it seems reasonable that the attempt frequency (typically of the order of a phonon frequency) is the same in both phases, we fixed $\tau_0$ for the PC to the value obtained for the SL/liquid state ($\tau_0 = 5.2 \times 10^{-15}$ s). For $T_{VF}$ we obtain 104 K (SL/liquid) and 100 K (PC). The glass temperatures deduced from an extrapolation of the VFT curves to $\tau = 100$ s are 144 K (SL/liquid) and 146 K (PC). Both are nearly identical, in agreement with the findings from DSC (section IIIA). However, as often found in supercooled liquids,[37] the glass temperatures obtained by dielectric spectroscopy are somewhat lower than the glass temperatures determined from DSC (section IIIA). We have also fitted the $\alpha$-relaxation time data of Fig. 6 with the recently proposed formula by Mauro et al.,[38] which is able to describe the data with equal quality as Eq. (1).

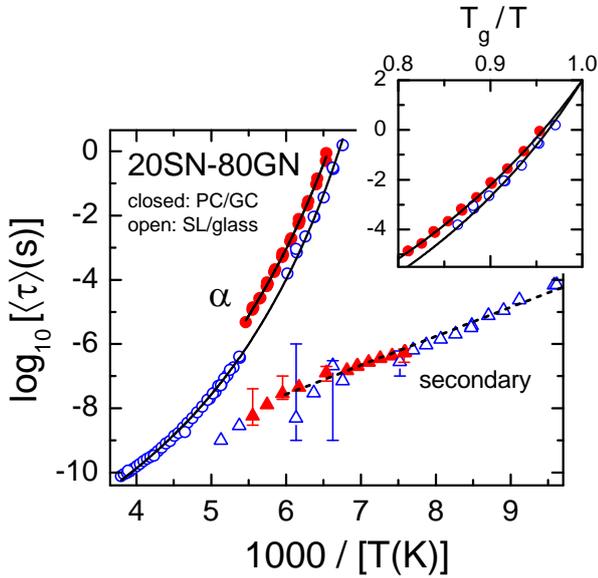

FIG. 6. Temperature-dependent average relaxation times[34] of the $\alpha$ (circles) and secondary relaxations (triangles) of 20SN-80GN. Closed symbols: results for the PC and glassy-crystal (GC) phases. Open symbols: Liquid, SL, and structural glass. The solid lines are fits using the VFT equation, Eq. (1). The dashed line is a fit with the Arrhenius formula. The inset shows an Angell plot of the $\alpha$-relaxation times of both phases close to $T_g$.

The strength parameter $D$ is used in the classification scheme for glass formers, introduced by Angell, to distinguish between strong and fragile glass formers.[36] While the relaxation times of fragile glass formers strongly deviate from Arrhenius behavior, these deviations are much weaker for strong glass formers. We obtain $D = 14.2$ (liquid and SL) and $D = 17.4$ (PC), i.e. in the PC state the relaxation dynamics has stronger characteristics than in the SL state. This finding is corroborated by the Angell plot[39] shown as inset of Fig. 6. In this $T_g$-scaled Arrhenius plot, the PC data points approach $T_g$ less steeply than for the SL. From the slopes of these curves at $T_g$ the fragility index $m$ can be determined[40] leading to $m = 47$ (PC) and $m = 62$ (SL). The fragility of glass formers was proposed to be linked to the form of the potential energy landscape in configuration space.[41] Within this framework, a higher fragility was assumed to be caused by a higher density of energy minima. Thus, the smaller fragility of the PC phase of 20SN-80GN can be rationalized by the fact that its lattice symmetry leads to a reduced density of minima compared to the supercooled-liquid phase, where the molecules, in addition to rotational, also possess translational degrees of freedom.

Nevertheless, with $m \approx 47$, the PC phase of 20SN-80GN still is rather fragile in comparison to most other PCs. Figure 7 shows an Angell plot containing relaxation-time data of a variety of PCs.[10,23,42,43] Most PCs exhibit strong or intermediate characteristics. However, 20SN-80GN, together with 60SN-40GN and Freon112, stands out by showing much more pronounced deviations from Arrhenius behavior. Here 60SN-40GN ($m = 62$)[23] is revealed to behave even more fragile than the PC phase of 20SN-80GN ($m = 47$). In Ref. 23, the relatively high fragility of the mixture 60SN-40GN was ascribed to a more complex energy landscape in comparison to PCs consisting of a single component. This increased complexity of the landscape was assumed to be caused by the additional substitutional disorder in this mixture and also by the fact that both components of the 60SN-40GN mixture are known to exist in different molecular conformations. The latter effect also explains the relatively fragile behavior of Freon112 ($m = 62$),[43] which has two conformers, leading to additional degrees of freedom. Interestingly, 20SN-80GN investigated in the present work has somewhat lower fragility than 60SN-40GN (cf. Fig. 7). This finding corroborates the suggested importance of the substitutional disorder for the energy landscape and, thus, for the fragility: 60SN-40GN is closer to a 50/50 molecule ratio which corresponds to maximum substitutional disorder while 20SN-80GN comes closer to a pure system.

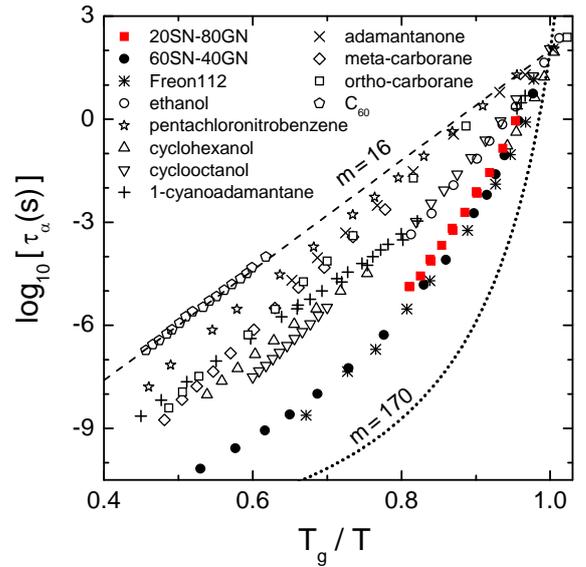

FIG. 7. Angell plot of the $\alpha$-relaxation times of various PCs.[10,23,42,43] The dashed line demonstrates maximally strong behavior; the dotted line shows an example for extremely high fragility. The data of the present work are shown as closed squares.

The triangles in Fig. 6 indicate the secondary-relaxation times $\tau_{sec}$ as obtained from fits as shown in Fig. 3. Obviously, within the error bars of the data, $\tau_{sec}$ of both phases agrees. Moreover, $\tau_{sec}(T)$ of Fig. 6 also reasonably agrees with the secondary relaxation times reported for 60SN-40GN and ascribed to a process denoted as $\gamma$ relaxation.[23] Thus, it seems unlikely that this relaxation corresponds to a so-called Johari-Goldstein $\beta$ relaxation,[27] which is assumed to be inherent to glassy matter and which usually is found to vary when the $\alpha$-relaxation parameters change.[44] Instead, just as reported for 60SN-40GN in Ref. 23, the "true" Johari-Goldstein $\beta$ relaxation of this material may be located between the detected secondary relaxation and the $\alpha$ relaxation. It may correspond to the additional relaxation suggested in section IIIB when discussing the spectrum at 132 K shown in Fig. 3(b). The dashed line in Fig. 7 is a fit using the Arrhenius law, $\tau = \tau_0 \exp[E/(k_B T)]$ ($\tau_0$: inverse attempt frequency, $E$: energy barrier). It leads to $E = 0.18$ eV and $\tau_0 = 1.0 \times 10^{-13}$ s, similar to the results on 60SN-40GN.[23] Interestingly, in supercooled and plastic crystalline ethanol also secondary relaxations were detected, whose relaxation times are nearly identical.[19,21] In Ref. 21 they were ascribed to coupled librational and intramolecular motions.

## IV. SUMMARY AND CONCLUSIONS

In the present work, we have shown that mixtures of succinonitrile and glutaronitrile belong to the rare examples of materials that can be prepared both in a supercooled-liquid and plastic-crystalline state. DSC investigations were used to provide a detailed phase diagram revealing that, depending on thermal history, the two disordered phases can be prepared for glutaronitrile concentrations between 70 and 90%. A thorough investigation of the 80% mixture by broadband dielectric spectroscopy has revealed that the glassy dynamics marked by its spectral behavior, relaxation time, and glass temperature is astonishingly similar in both disordered phases of this material. This finding is in good accord with those in ethanol, the only material where the glassy dynamics of both phases has been thoroughly characterized until now.[10,18,19,20,21] This implies that the conclusions drawn from the results in ethanol, namely that reorientational motions play an important and often underestimated role for the glass transition, indeed seem to be generally valid also for other types of glass formers.

Moreover, we have demonstrated that plastic-crystalline 20SN-80GN belongs to the few cases of PCs that show relatively fragile dynamics. This finding and the different fragilities of the PC and SL phases of 20SN-80GN and of plastic crystalline 60SN-40GN can be well understood within an energy-landscape-related framework used to explain the fragilities of glass formers.[41]

## ACKNOWLEDGMENTS

We thank C. A. Angell for stimulating discussions. This work was partly supported by the Deutsche Forschungsgemeinschaft via Research Unit FOR 1394.